\documentclass[%
 reprint,
 amsmath,amssymb,
 aps,onecolumn,12pt]{revtex4-1}
\usepackage{subfig}
\usepackage{graphicx}
\usepackage{float}
\usepackage{dcolumn}
\usepackage{bm}
\usepackage{sistyle}
\usepackage{braket}
\usepackage{lipsum}
\usepackage{titlesec}
\usepackage[]{subfig}
\usepackage{color}
\begin{document}
\title{DFT+U Investigation of magnetocrystalline anisotropy of Mn-doped transition-metal dichalcogenides monolayers}

\author{Adlen Smiri}
\affiliation{Facult\'e des Sciences de Bizerte, Laboratoire de Physique des Mat\'eriaux: Structure et Propri\'et\'es,\\
Universit\'e  de Carthage, 7021 Jarzouna, Tunisia}%
\affiliation{%
LPCNO, Universit\'e F\'ed\'erale de Toulouse Midi-Pyr\'en\'ees,\\
INSA-CNRS-UPS, 135 Avenue de Rangueil, 31077 Toulouse, France
}%
\author{Sihem Jaziri}
\affiliation{Facult\'e des Sciences de Bizerte, Laboratoire de Physique des Mat\'eriaux: Structure et Propri\'et\'es,\\
Universit\'e  de Carthage, 7021 Jarzouna, Tunisia
}%
\affiliation{
Facult\'e des Sciences de Tunis, Laboratoire de Physique de la Mati\'ere Condens\'ee,\\
D\'epartement de Physique, Universit\'e  Tunis el Manar, Campus Universitaire 2092 Tunis, Tunisia
}%
\author{ \small {Samir Lounis}}
\affiliation{{\footnotesize Peter Gr\"unberg Institut and Institute for
Advanced Simulation, Forschungszentrum J\"ulich and JARA, 52425 J\"ulich, Germany
}}%
 \author{Iann C. Gerber}
\email{igerber@insa-toulouse.fr}
\affiliation{%
LPCNO, Universit\'e F\'ed\'erale de Toulouse Midi-Pyr\'en\'ees,\\
INSA-CNRS-UPS, 135 Avenue de Rangueil, 31077 Toulouse, France
}%

%
\begin{abstract}
Doped transition-metal dichalcogenides monolayers exhibit exciting magnetic properties for the benefit of two-dimensional spintronic devices. Using density functional theory (DFT) incorporating Hubbard-type of correction (DFT$+U$) to account for the electronic correlation, we study the magnetocrystalline anisotropy energy (MAE) characterizing Mn-doped MS$_2$ (M=Mo, W) monolayers. A single isolated Mn dopant exhibits a large perpendicular magnetic anisotropy of 35 meV (8 meV) in the case of Mn-doped WS$_2$ (MoS$_2$) monolayer. This value originates from the Mn in-plane orbitals degeneracy lifting due to the spin-orbit coupling. In pairwise doping,  the magnetization easy axis changes to the in-plane direction with a weak MAE compared to single Mn doping. Our results suggest that diluted Mn-doped MS$_2$ monolayers, where the Mn dopants are well separated, could potentially be a candidate for the realization of ultimate nanomagnet units.
\end{abstract}
\maketitle
\section{Introduction}
Inducing magnetism in nonmagnetic semi-conductor atomically thin monolayers (MLs) is a current field of investigation in material science to reach applications in storage and quantum spin processing. Interestingly, only  very few experimental works on doping MoS$_2$ MLs with other transition metals~\cite{ref017, ref14} are available, when most of the reported studies are theoretical ones~\cite{ref6,ref7,ref9,ref060,ref047,ref10,ref8}. It has been shown that doping can induce strong ferromagnetism~\cite{ref6,ref7,ref8,ref9,ref10,ref14,ref15,ref015,ref016,ref060} and large magnetic anisotropy that corresponds to direction-dependent magnetism~\cite{ref039,ref040,ref044,ref047,ref048} in two-dimensional (2D) materials. MoS$_2$ ML appears to be an emblematic case in the family of transition metal dichalcogenide semiconductors (TMDs) since it possesses peculiar physical properties~\cite{ref010,ref011,ref012,ref013,ref014}. 
It is characterized by a robust excitonic binding energies of hundreds of meV which suggests several potential applications such as laser or light-emitting diodes fields~\cite{ref017,ref018,ref019,ref020}. Furthermore, MoS$_2$ ML has specific transport properties~\cite{ref049,ref050,ref051,ref054,ref051,ref052,ref053}. In particular, this 2D semiconductor has high low-temperature electron mobility (up to 1000 cm$^2$/Vs)~\cite{ref049,ref050,ref051,ref054} and low-power dissipation~\cite{ref051,ref052,ref053} which makes it a good candidate to build transistors~\cite{ref051,ref052,ref053,ref055}. Thus inducing magnetism in this type of well-featured materials  is of first importance to address spintronic applications~\cite{ref7,ref8,ref9,ref14,ref017}. The substitution of Mo atoms in $2H$-MoS$_2$ ML by Mn ones is one way to realize promising magnetic 2D-TMDs candidates~\cite{ref7,ref8,ref9,ref14,ref017}.  From a thermodynamical point of view, exchanging Mn at Mo sites is found energetically favorable under S-rich regime~\cite{ref8}. Moreover, it has been shown that Mn dopants clustering within MoS$_2$ ML appears to be a more stable configuration than the well-dispersed Mn dopants case~\cite{ref10}. Experimentally, doped Mn-MoS$_2$ ML samples have been realized either by using vapor phase deposition techniques for low doping concentration~\cite{ref010}, or by a hydrothermal method to reach higher doping concentration~\cite{ref14}. Theoretical studies have shown that doping MoS$_2$ ML with Mn promotes a strong ferromagnetism with high Curie temperatures~\cite{ref7,ref8,ref9,ref14,ref017}.

In addition to the desired goal of high Curie temperature in 2D-TMDs, it is highly desirable to have a large magnetic anisotropy energy (MAE), which defines the energy barrier preserving the magnetic moments in a preferential direction against thermal fluctuations at room temperature~\cite{Khajetoorians976,PhysRevLett.101.137201,ref5}. In order to achieve high density magnetic data storage devices, the magnetic anisotropy induced by single adatoms or single atoms is of great interest~\cite{ibanez,rau,donati,ref047,ref048}. For substitutional Mn-doped MoS$_2$ ML, to our knowledge most of the studies focused on the ferromagnetism aspect without addressing its MAE. The only works dealing with MAE estimates in doped TMDs, have considered (i) magnetic adatoms (Mn or Fe) on MoS$_2$ ML containing S di-vacancies~\cite{ref047} which reach MAE values of few meVs with a preferential in-plane magnetization for Mn adatom and a preferential out-of-plane magnetization for a single adsorbed Fe; (ii) the doping of WS$_2$ ML by substitutional Co and Fe atoms that can achieve large perpendicular magnetic anisotropy of few tenths of meV~\cite{ref048}.\\

\begin{figure}
  \begin{center}
      \includegraphics[width=0.75\textwidth]{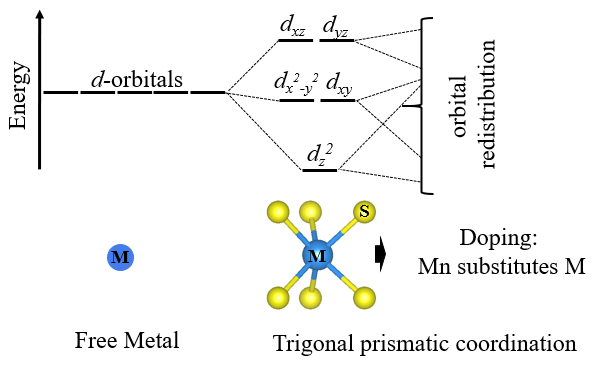} 
\end{center}
\caption{Schematic diagram explaining the M (M = Mo or W) $d$-orbitals splitting under crystal field of the trigonal prismatic symmetry. The $d$ orbitals of the isolated M atom (left) split into three categories (middle) for MS$_2$ MLs. The $d$ orbitals are expected to show, locally, further splitting when the Mn dopant takes the place of the M atom.
\label{fig:fig(trigonal)}}
\end{figure}

It is well known that in pristine \textit{2H}-MS$_2$ ML (with M=Mo, W), under a trigonal prismatic coordination, the \textit{d}-M orbitals are split in three categories~\cite{Li,ref12} in the associated point group $D_{3h}$ (figure~\ref{fig:fig(trigonal)}). The M lowest energy orbital corresponds to \textit{d$_{z^2}$} which reflects its weak coupling with the surrounding atoms. The M intermediate energy orbitals \textit{d$_{xy}$} and \textit{d$_{x^2-y^2}$} are coupled to each other, when the highest energy orbitals \textit{d$_{xz}$} and \textit{d$_{yz}$} are strongly coupled with 3\textit{p} S atom. As a consequence, those orbitals dominate the conduction and valence band edges characters in MS$_2$ ML~\cite{ref11,ref12,ref13}, and are the origin of strong spin-orbit coupling (SOC) in the valence band~\cite{ref059,Zhu} in K-point of the Brillouin zone, up to 426 and 150 meV for WS$_2$ and MoS$_2$ MLs, respectively. Any substitution of a M atom by a dopant, such as a Mn atom, should lead to a local orbital redistribution with respect to the energy due to crystal field effects~\cite{ref6,ref7,ref8}, with spin splitting. The dopant orbital magnetic moment direction is thus settled by both SOC and the crystal field effect, possibly inducing MAE~\cite{ref058,ref059}. Understanding and possibly tuning the MAE of Mn-doped MS$_2$ ML appear crucial to prospect the potential of these materials in information storage devices.\\

In this work, density functional theory (DFT) corrected by a Hubbard term (DFT+$U$) to account for strong electronic correlation in $3d$ orbitals, is considered to study the possible magnetic anisotropy in the Mn doped MS$_2$ ML systems. The importance of such correlations were previously recognized in bulk $3d$ Ni and Fe atoms, where their account led to MAEs with a better agreement with experiments~\cite{yang,klatyk}, since these effects can trigger large orbital magnetic moments~\cite{yang}. Another possibility as in Ref.~\cite{Li}, would be to use hybrid functionals, to provide a better localization of the $3d$ orbitals, as proposed to obtain reliable spin orbit splitting of defect states at the origin of the MAE. However the computational cost associated to this task remains very too large. Hence, taking into account the $U$-dependence, as done in the present work, is necessary to establish a complete study of the magnetic anisotropy of Mn-MS$_2$ doped MLs.\\

Using DFT+$U$ method, we have estimated and compared the MAE in single and pairwise Mn-doped MS$_2$ MLs. In particular, we have first investigated the MAE induced by a single isolated Mn dopant, representative of a moderate dopant's concentration around 4\%. Then, in view of the preferential clustering of substitutional Mn atoms in MS$_2$ ML, we have studied the MAE's sensitivity to the separation distance between Mn atoms. To this end, various configurations have been constructed by placing two Mn atoms in M sites of a zigzag or/and armchair patterns with different Mn-Mn separations. In addition, for relevant doping cases, through the analysis of dopant SOC we have investigated and rationalized the MAE's origin.
\section{Methods and Computational details}
Spin-polarized DFT as implemented in Vienna \textit{ab initio} simulation package (VASP) was used in this work~\cite{ref1,ref2}. The core potential was approximated by the projected augmented wave (PAW) scheme~\cite{ref3,Kresse:1999a}. Perdew-Burke-Ernzerhof (PBE) formulation of generalized gradient approximation (GGA) was applied to describe the exchange-correlation interaction~\cite{ref4}. In addition, a Hubbard U correction~\cite{DFTU} was adopted in order to better describe the localization of 3\textit{d} Mn orbitals. The criteria of atom force convergence, used for all structure relaxations, was fixed to 0.02 eV/$\angstrom$, with an energy cutoff of 400 eV. In order to model the geometry of Mn-doped MS$_2$ MLs, supercells of size 5$\times$5$\times$1 and 7$\times$7$\times$1 were considered, with a 20-$\angstrom$-thick vacuum region to separate adjacent MLs. The initialized lattice parameters are equal to 3.18 $\angstrom$ for WS$_2$ ML and 3.20 $\angstrom$ for MoS$_2$ ML. To calculate the MAE, a 4$\times$4$\times$1, 7$\times$7$\times$1 and 9$\times$9$\times$1 $\Gamma$-centered Monkhorst-Pack grid with the tetrahedron smearing method of Bl\"ochl~\cite{bloechl:prb:94_b} were tested to obtain accurate results. For density of state calculations, the $k$-points grid has been increased to 12$\times$12$\times$1 and the smearing method was switched to the Gaussian type with 0.02 eV width.\\

The magnetic anisotropy energy of Mn-doped MS$_2$ is determined by performing a fully self-consistent SOC and non-collinear calculations with different orientations of the magnetic moments of Mn impurities. In particular, we calculate the total energies $E_{x}$ and $E_{z}$ when the magnetization direction is parallel to $x$- and $z$-axis, respectively. Therefore, the MAE is given by
\begin{equation}
\label{eqdft}
\text{MAE}=E_{x}-E_{z},
\end{equation}
 where $x,y,z$ represent the crystalline axes with the $z$-axis perpendicular to the ML plane. Positive MAE stands for preferential out-of-plane magnetization corresponding to perpendicular magnetic anisotropy (PMA), while negative MAE stands for a preferential in-plane magnetization which is known as in-plane magnetic anisotropy (IMA). Furthermore, in order to elucidate the origin of MAE, spin-orbit energy is calculated, using a scalar relativistic approximation~\cite{koelling} as implemented in VASP code~\cite{hafner}, through the following expression,
 \begin{equation}
\label{eqsoc}
\displaystyle E_{\text{SOC}}=\langle \frac{\hbar^2}{2m^2c^2}\frac{1}{r}\frac{dV(r)}{dr}\bm{\hat{L}.\hat{S}}\rangle,
\end{equation}
Here, $\bm{\hat{L}}$ and $\bm{\hat{S}}$ denote the orbital and spin  angular momentum operators, respectively, when $V(r)$ is the spherical part of the effective potential within the PAW sphere.\\

\section{Results and discussion}
\subsection{Magnetic anisotropy of an isolated Mn atom in MoS$_2$ ML}
\subsubsection{$U$-dependence of magnetic anisotropy}
\begin{figure}[!h]
  \begin{center}
     \subfloat[]{
      \includegraphics[width=0.65\textwidth]{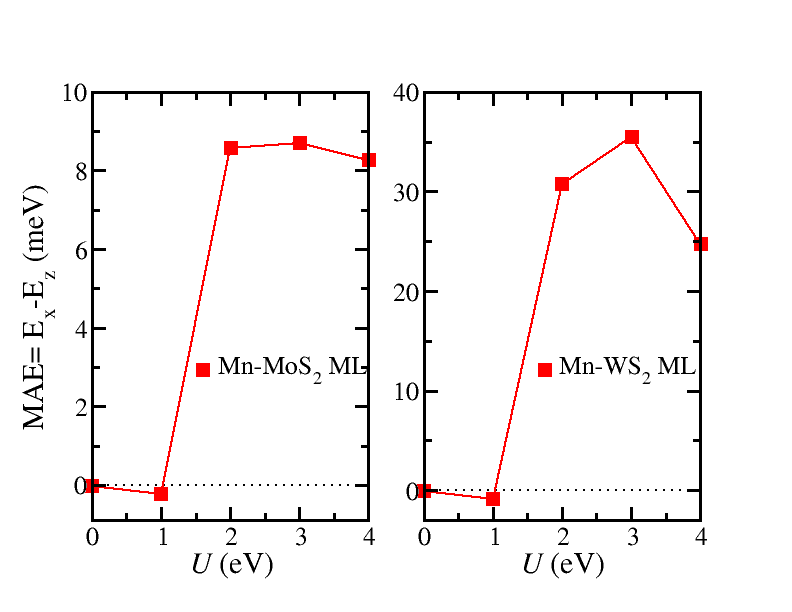} 
    \label{fig(c0)}                     
                         } 
\qquad
     \subfloat[]{
      \includegraphics[width=0.6\textwidth]{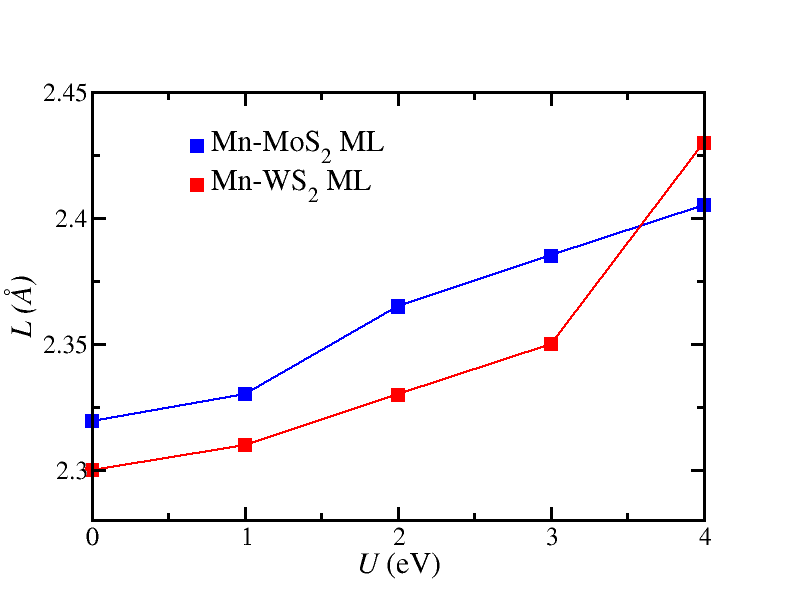} 
     \label{fig:fig(c1)}                    
                         } 
\end{center}
\caption{(a) The magnetic anisotropy energy (MAE) of Mn-MS$_2$ doped MLs as a function of the Hubbard parameter $U$. (b) The length $L$ of the bonds formed between Mn dopant and the first nearest neighboring S host atoms as a function of $U$.
\label{ffig(c0)}}
\end{figure}

To study the electronic correlation effects on the magnetic anisotropy of Mn-MS$_2$ doped MLs, the calculated MAEs with respect to $U$-parameter are shown in figure~(\ref{fig(c0)}). Note here that for each Hubbard term $U$, the geometrical optimization is repeated. In the absence of the correlation correction, the easy axis of magnetization is oriented along $x$ which indicates an IMA for the two systems. Here, the MAE value of Mn-WS$_2$ doped ML is in good agreement with that of Ref.~\cite{ref048}. We find that without $U$, the magnitude of MAEs are almost the same for both investigated systems. However, by introducing the correlation effects, a change in the magnetization easy axis occurs at $U = 2$ eV resulting in an energetically favored PMA states, with MAEs much larger for Mn-WS$_2$ ML than for Mn-MoS$_2$ ML. A similar change of MAE sign under $U$-parameter variation was found for Fe and Ni atoms~\cite{yang}. We note that Mn-WS$_2$ ML MAE is larger than Mn-MoS$_2$ ML one.\\

In particular, for $U>$ 1 eV, Mn-MoS$_2$ doped ML exhibits a large MAE of 8 meV compared to the results of Cong \textit{et al.},~\cite{ref047} where the largest MAE value was found to be 1.3 meV for Mn adatom implanted in a di-sulfur vacancy in MoS$_2$ ML. Higher MAEs are found in the case of Mn-WS$_2$ doped ML for the same range of $U$. Indeed, the MAE reaches 30, 35 and 24 meV for the Hubbard parameter of 2, 3 and 4 eV, respectively. A thermal stability at room temperature of the magnetic anisotropy can be envisaged in the case of Mn doped-WS$_2$ ML.\\
\subsubsection{Influence of crystal structure  on the magnetic anisotropy}
\begin{figure}[!h]
  \begin{center}
     \subfloat[$U=$ 0 eV]{
      \includegraphics[width=0.45\textwidth]{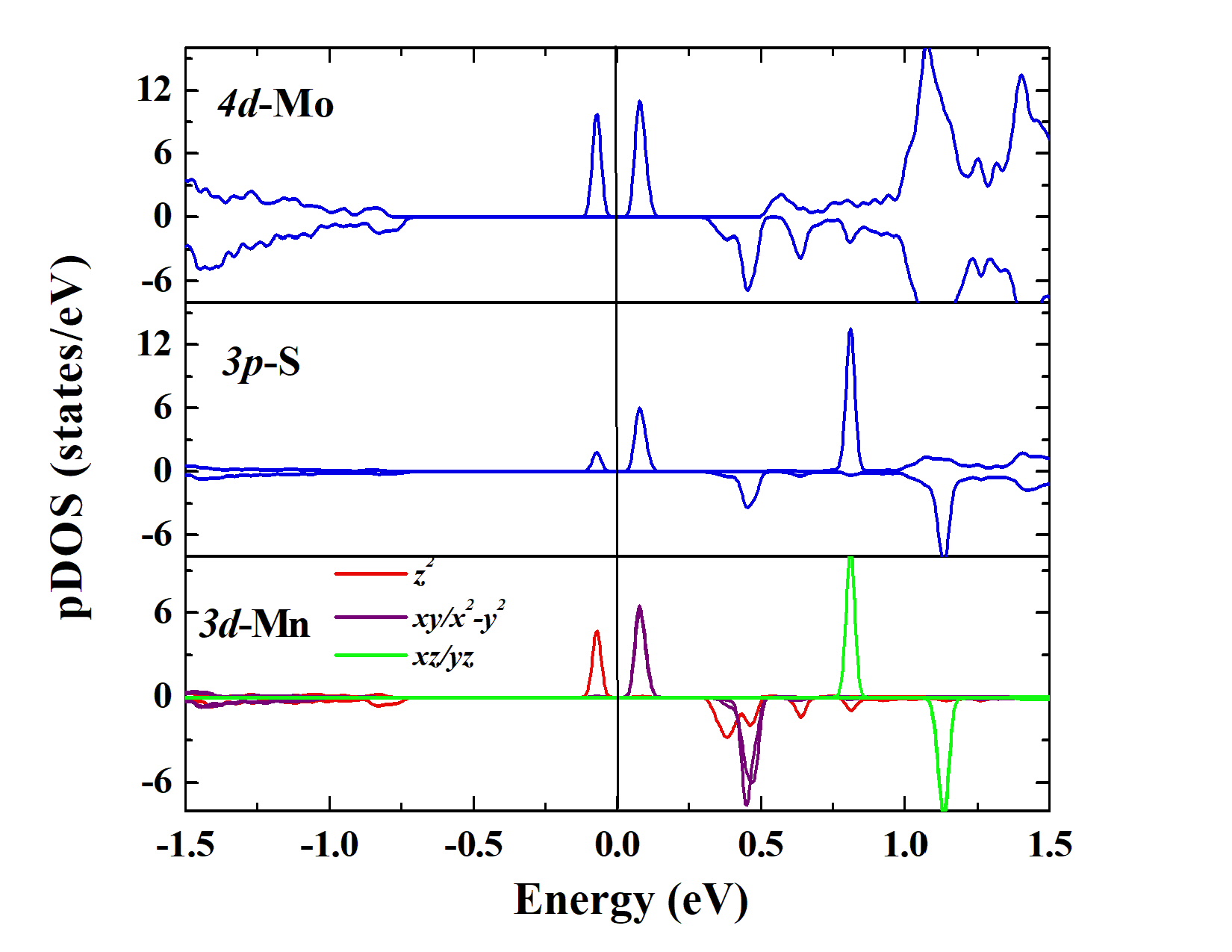} 
      \label{fig(c31)}
                         }  
      \subfloat[$U=$ 3 eV]{
      \includegraphics[width=0.45\textwidth]{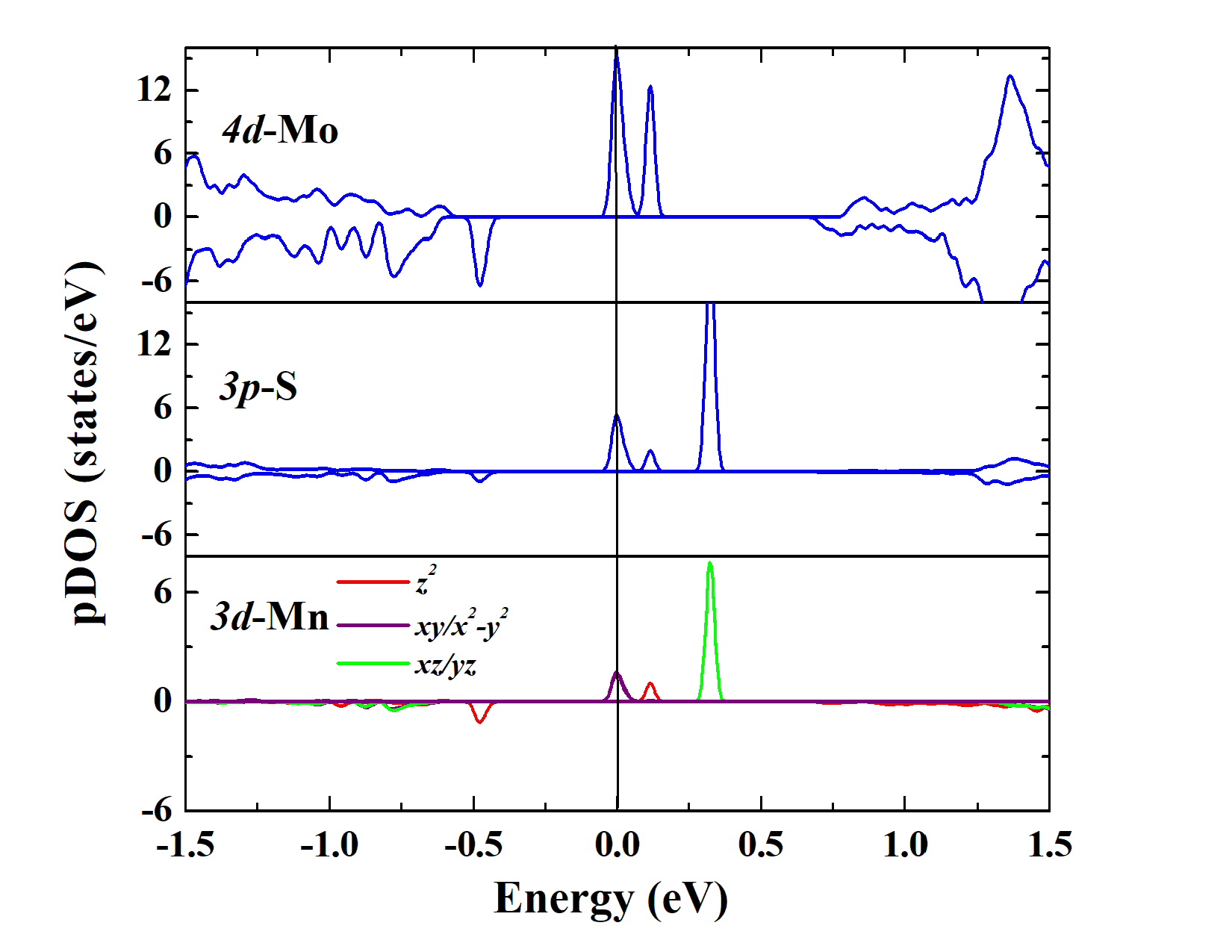}
      \label{fig(c32)}
                }
    \qquad
         \subfloat[$U=$ 0 eV]{
      \includegraphics[width=0.45\textwidth]{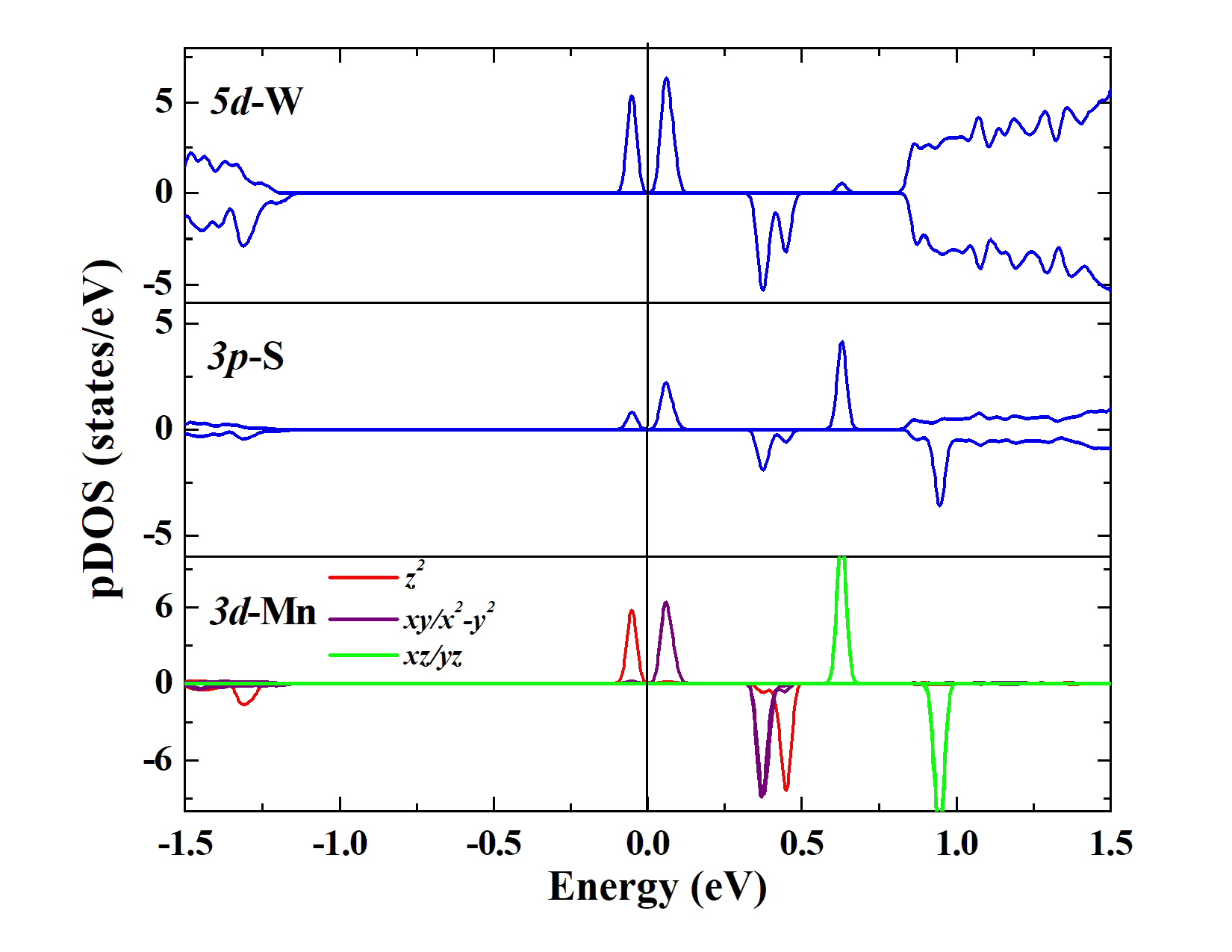} 
      \label{fig(c33)}
                         }  
      \subfloat[$U=$ 3 eV]{
      \includegraphics[width=0.45\textwidth]{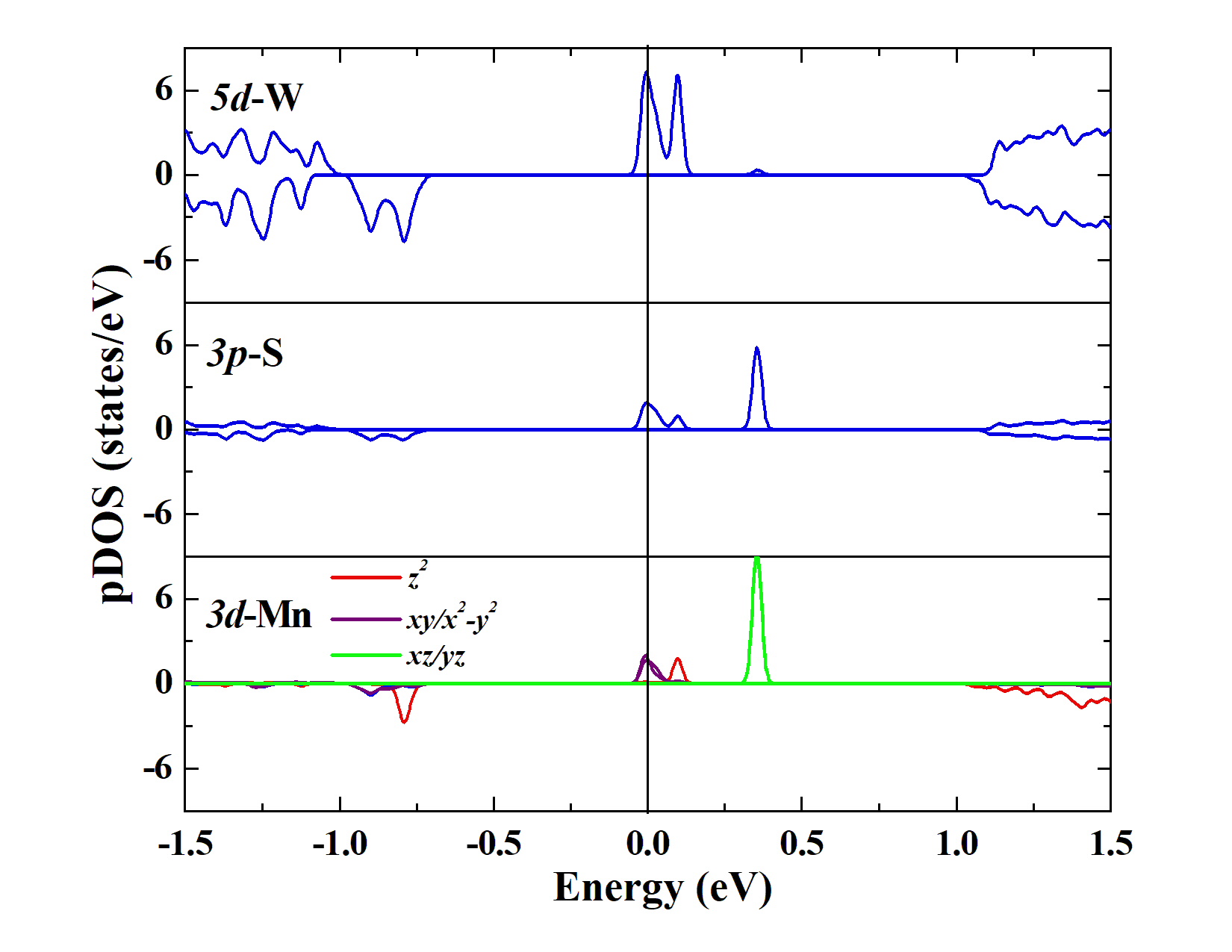}
      \label{fig(c34)}
                         }  
\end{center}
\caption{The DFT (a and c) and the DFT$+U$ (b and d) (for $U=$ 3 eV) pDOS of $3d$ states of Mn dopant and $d$ and $4p$ states of its neighboring M and S atoms, respectively.
\label{fig:fig(c3)}}
\end{figure}

Owing to the Hubbard $U$-parameter, the equilibrium atomic positions can vary and thus impact the electronic structures  Hence, a connection between the MAE variation and the structural modifications, both under $U$-parameter, is possible. To reveal the effect of correlations on the local crystal structure, the S-Mn bond length ($L$) is plotted in figure~(\ref{fig:fig(c1)}) as a function of $U$-parameter upon geometry optimization. Interestingly when $U$ is increased, $L$ becomes larger too.
In the meantime, the on-site Coulomb correlation enhances the magnetic anisotropy magnitude. The enhancement of MAE seems to be correlated to the increase in the S-Mn bond lengths suggesting a common electronic origin.
In term of electronic structure, the $L$ increase results in a decrease of the overlapping between $3p$ S and $3d$ Mn orbitals. To explore the orbital distribution and hybridization in Mn doped MS$_2$ MLs, the projected density of states (pDOS) for DFT and DFT$+U$ are shown in figure~(\ref{fig:fig(c3)}). From DFT, the $3d$ Mn orbital energy level split into three different levels under crystal field effect of the trigonal prismatic coordination: a twofold degenerate level containing the out-of-plane orbitals ($d_{xz}$, $d_{yz}$) with the highest energies among $3d$ Mn levels; a twofold degenerate level containing the in-plane orbitals ($d_{xy}$, $d_{x^2-y^2}$) with an intermediate energies and non-degenerate level containing the perpendicular orbital ($d_{z^2}$) with the lowest energies. The $d_{xz}$ and $d_{yz}$ level has the highest energy because they are oriented toward the chalcogen atoms as indicated in the figure~(\ref{fig:fig(trigonal)}). Hence, $d_{xz}$ and $d_{yz}$ orbitals have a greater overlapping with $3p$ S orbitals than the rest of Mn orbitals that are lower in energy. Interestingly, once the correlation effect included, $d_{xz}$ and $d_{yz}$ level starts to shift toward lower energies  which explains the increase of the S-Mn bond lengths.\\

\begin{figure}[!h]
  \begin{center}
       \subfloat[Mn-MoS$_2$ doped ML]{
      \includegraphics[width=0.5\textwidth]{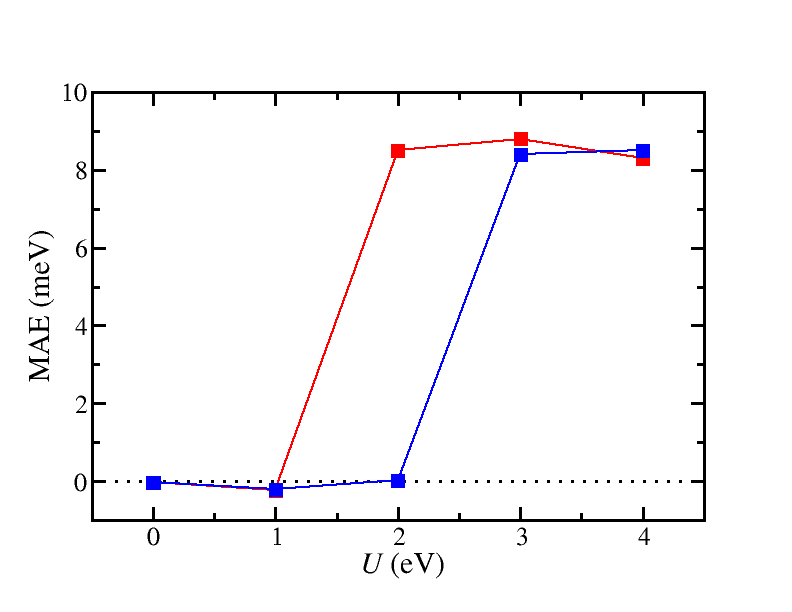} 
      \label{fig(c21)}
                         }
            \subfloat[Mn-WS$_2$ doped ML]{
      \includegraphics[width=0.5\textwidth]{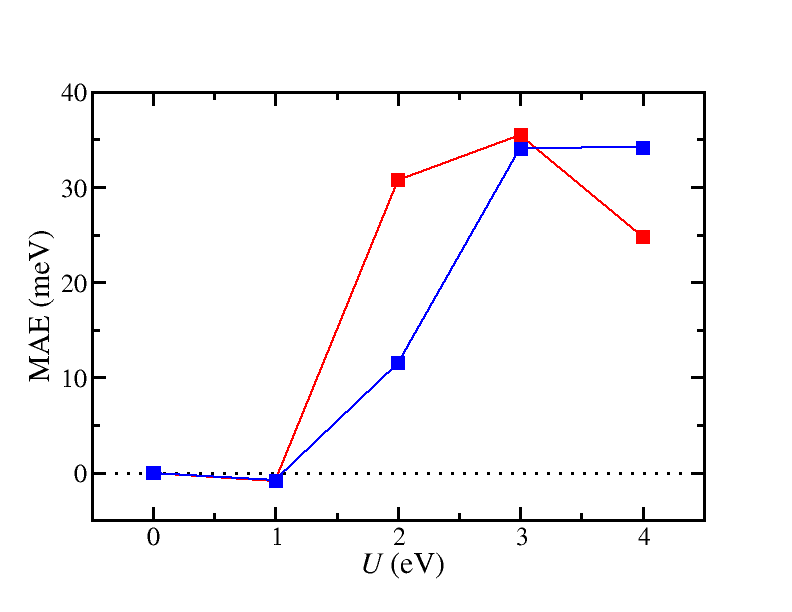} 
      \label{fig(c22)}
                         }                 
\end{center}
\caption{The MAE values with respect to the values of the correlation parameter $U$ for $U$-dependent (red lines) and -independent (blue lines) crystal structure of (a) Mn-MoS$_2$ doped ML and (b) Mn-WS$_2$ doped ML.
\label{fig:fig(c2)}}
\end{figure}

\begin{figure} [H]
  \begin{center}
     \subfloat[Mn-doped MoS$_{2}$ ML]{
      \includegraphics[width=0.55\textwidth]{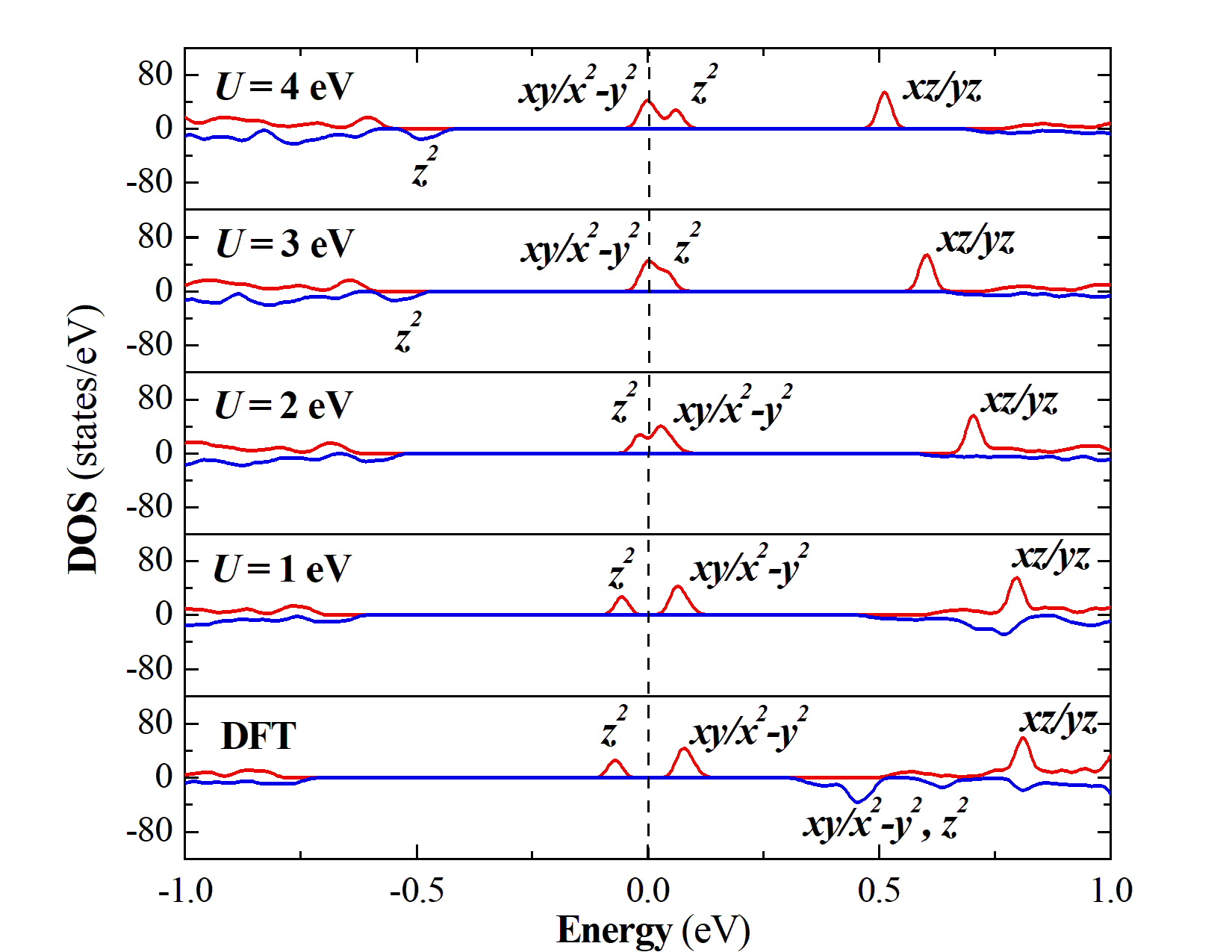} 
      \label{fig(c23)}
                         }  
      \subfloat[Mn-doped MoS$_{2}$ ML]{
      \includegraphics[width=0.55\textwidth]{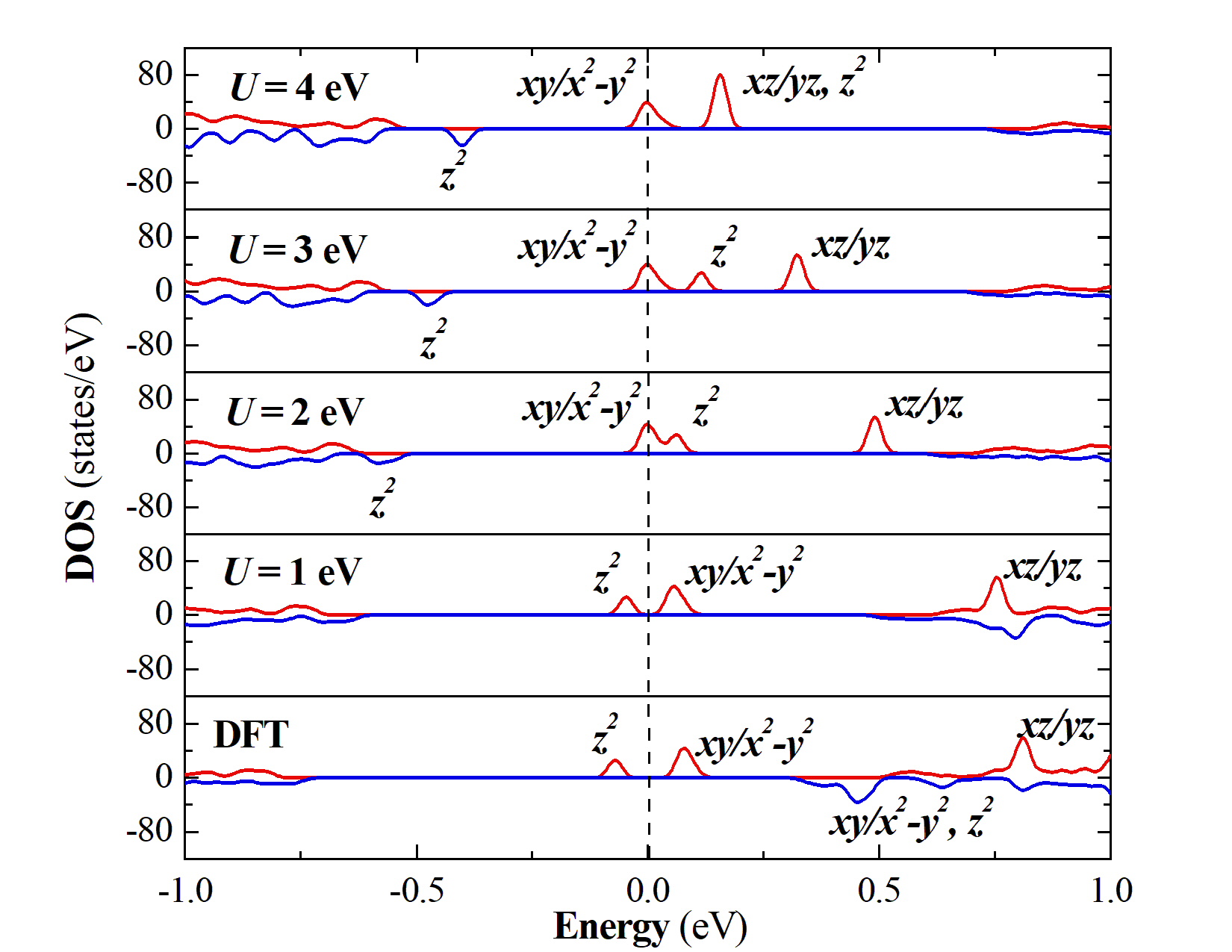}
      \label{fig(c24)}
                         }
      \quad
     \subfloat[Mn-doped WS$_{2}$ ML]{
      \includegraphics[width=0.55\textwidth]{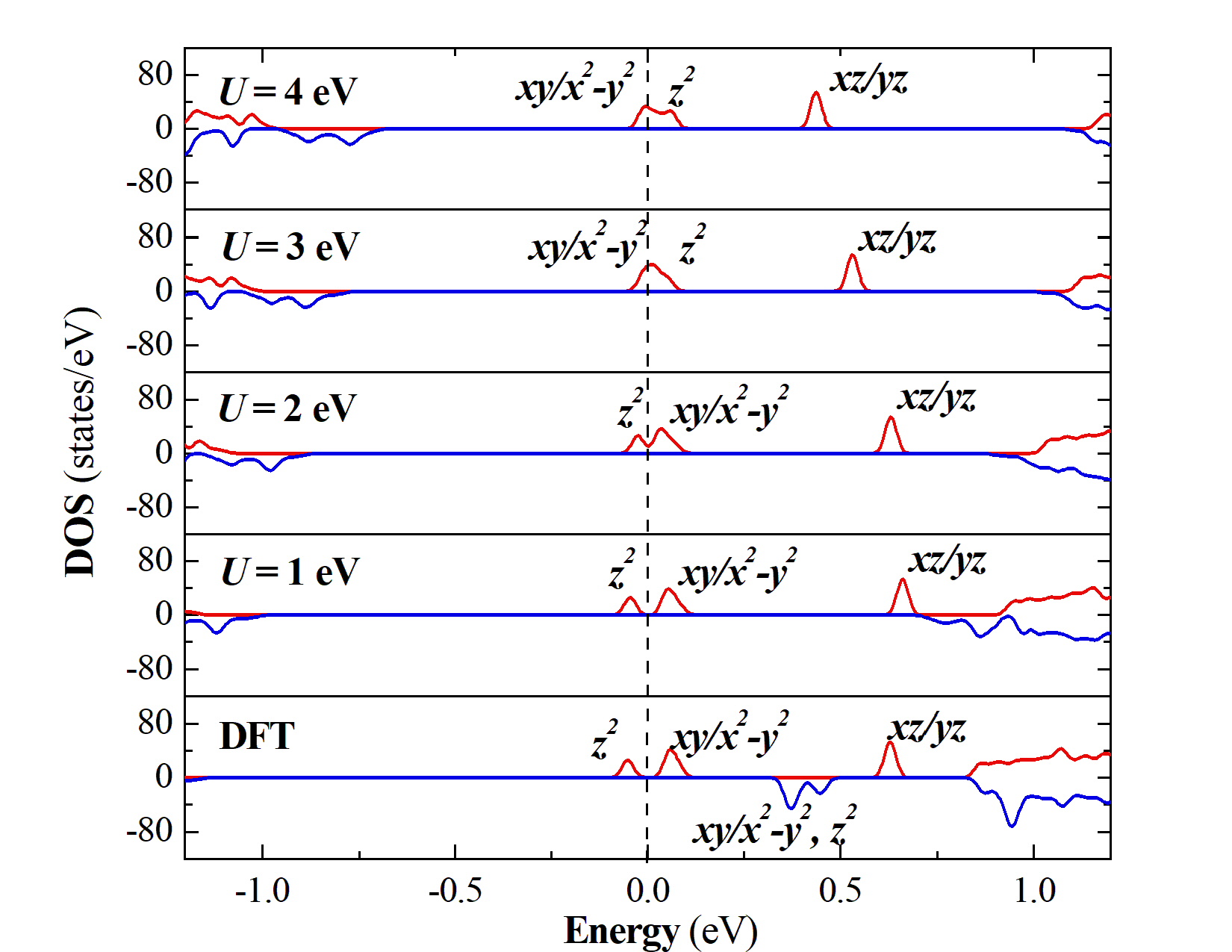} 
      \label{fig(c25)}
                         }  
      \subfloat[Mn-doped WS$_{2}$ ML]{
      \includegraphics[width=0.55\textwidth]{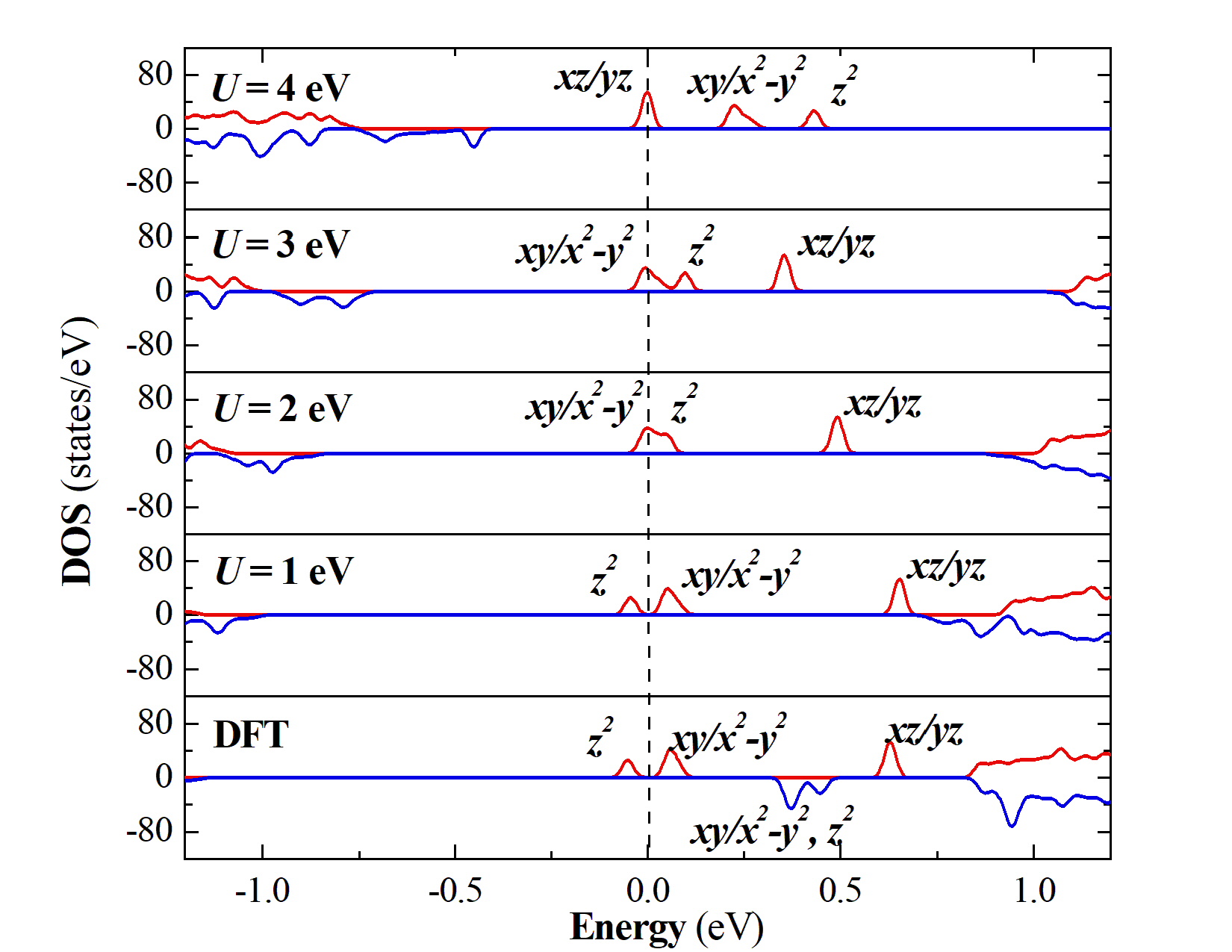}
      \label{fig(c26)}
                         }                       
\end{center}
\caption{The total electronic density of states (DOS) of $U=$ 0 to 4 eV for (b) $U$-dependent and (c) $U$-independent crystal structure of Mn-doped MoS$_{2}$ ML and Mn-doped WS$_{2}$ ML.
\label{fig:fig(c20)}}
\end{figure}

Furthermore, the decrease of the hybridization, under $U$ variation, occurs for all Mn $3d$ orbitals and not only $d_{xz}$ and $d_{yz}$ orbitals. The Mn orbitals ($d_{xy}$, $d_{x^2-y^2}$) and $d_{z^2}$ mainly hybridize with the $d$ orbitals of the metal M, see figure~(\ref{fig:fig(c3)}). For $U \in$ [0;3] for M=W and $U\in$ [0;4] for M=Mo (figure~(\ref{fig(c24)})); the in-plane $d_{xy}$ and $d_{x^2-y^2}$ levels slightly shift in energy until they cross the Fermi level and exceed the spin-up $d_{z^2}$ level. In the meantime, a spin-down $d_{z^2}$ state appears close to the valence band which maintains the same energetic order of $3d$ orbitals.\\
To further elucidate the MAE connection with local modifications of the geometries, we recalculate the MAE with a fixed lattice parameters of about 3.2$\angstrom$ obtained upon a DFT geometry optimization (case (II)) for MoS$_2$ ML. The results are plotted in figure~(\ref{fig:fig(c2)}) and compared to the ones \textcolor{red}{obtained} after complete lattice parameter optimization for each $U$ as already discussed (case (I)). When comparing the MAEs variations with $U$ of cases (I) and (II), the major differences appear at $U=$ 2 eV for both systems and at $U=$ 4 eV for Mn-WS$_2$ ML. In particular, the MAEs of case (I) shows a larger (smaller) values at $U=$ 2 (4) eV with respect to MAEs of case (II) for MoS$_2$ (WS$_2$) MLs. Therefore, the dependence of the crystal structure on $U$ can potentially be critical in determining the MAE based on the value of $U$ itself.\\
We try now to elucidate the electronic origin of the changes and similarities between MAEs obtained in the cases (I) and (II) with or without geometry optimizations. In particular, for each $U$ parameter, the DOS is represented for $U$-dependent and -independent crystal structures in figure~(\ref{fig:fig(c20)}). For $U<$ 2eV, the DOS of case (I) and (II) show comparable features which explain the coincidence between the corresponding MAEs. In particular, the Fermi level lays between the $d_{z^2}$- and $d_{xy}$/$d_{x^2-y^2}$-orbitals. For $U=$ 2 eV, the latter electronic properties are obtained in the case (II), whereas, the Fermi level crosses the spin-up bands of the two-fold degenerated $d_{xy}$/$d_{x^2-y^2}$ orbitals in the case (I). This explains the sudden increase of MAE at $U=$ 2 eV when the $U$-dependence of the crystal structure is considered. For $U>$ 2 eV, a semi-metallic behaviour is observed for both cases. For $U=$ 4 eV, the difference between the MAEs of the cases (I) and (II) in Mn-WS$_2$ system,  originates from the orbital type present at the Fermi level. Furthermore, by comparing the $3d$ energy positions in $U$-dependent and -independent DOS (figure~(\ref{fig:fig(c20)})), it appears that the on-site Coulomb correlation weakens the hybridization between Mn and its environment, Regarding MAE variations, the bond-length increase due to $U$ geometrical optimization is important only when $U=$ 2 eV for the two MLs and also for $U=$ 4 eV in the case of Mn-WS$_2$ doped ML.\\
\subsubsection{Connection between magnetic anisotropy and orbital magnetic moments}
\begin{figure} 
  \begin{center}
     \subfloat[]{
      \includegraphics[width=0.6\textwidth]{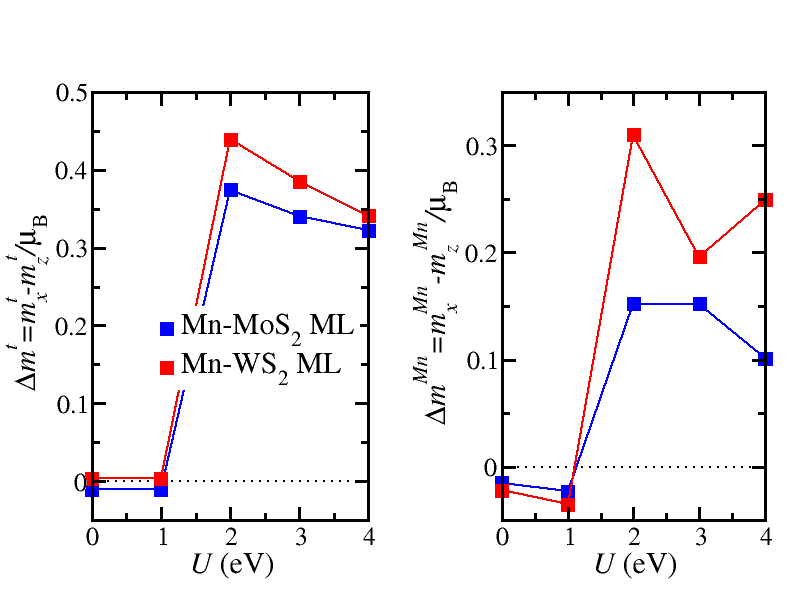} 
      \label{fig(cc0)}
                         }
    \quad                     
   \subfloat[]{
      \includegraphics[width=0.6\textwidth]{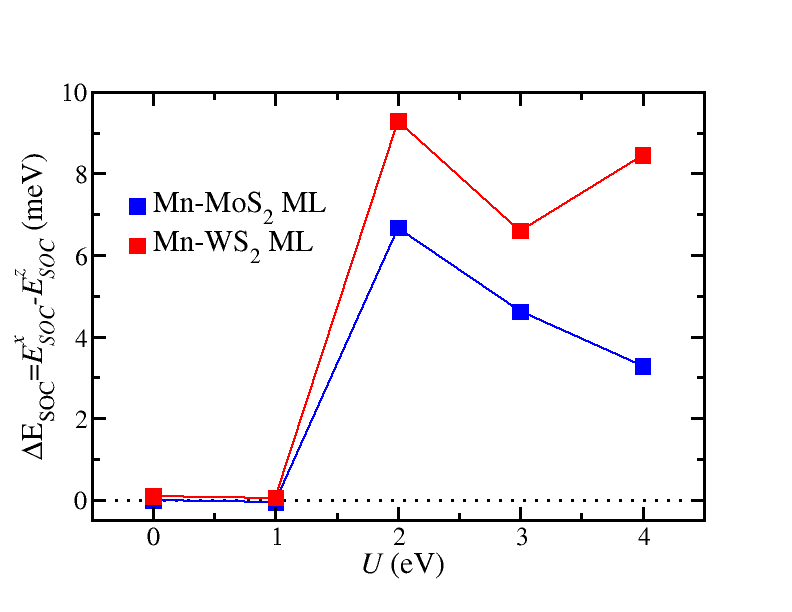} 
      \label{fig(so)}
                         }  
\end{center}
\caption{(a) The total (left) and local (right) orbital magnetic anisotropy (OMA) of Mn-MS$_2$ doped MLs as a functions of Hubbard parameter $U$. (b) The spin-orbit anisotropy energy of Mn-MS$_2$ doped MLs as a functions of Hubbard parameter $U$.
\label{fig(cc10)}}
\end{figure}

\begin{table}[H]
  \centering
  \resizebox{0.6\textwidth}{!}{
  \begin{tabular}{cc|ccc|cccccccccccccc}
  \hline
  \hline
    $U$ (eV) & & \multicolumn{3}{c|}{Mn-MoS$_2$ doped ML}  &&\multicolumn{3}{c}{Mn-MoS$_2$ doped ML} \\
  & & $m_z^t (\mu_B$) && $m_x^t (\mu_B$)&& $m_z^t (\mu_B$) && $m_x^t (\mu_B$)\\
 \hline
 0 && -0.00              &&-0.01   &&  -0.03     &&-0.03\\

 1 &&-0.00               && -0.01  &&  -0.03     &&-0.02\\ 

 2 && -0.38              && -0.01   &&  -0.46     &&-0.02\\

 3 && -0.36              &&-0.02   &&  -0.41     &&-0.03\\
   
 4 && -0.34              &&-0.02   &&  -0.32     &&0.02\\  
   \hline
   \hline
\end{tabular}
}
  \caption{The orbital magnetic moments and the atomic contributions of Mn dopant to and spin-orbit energy for in plane and out of plane magnetization orientation.}
  \label{tab(c0)}
\end{table}
The relationship between magnetic anisotropy and orbital moment anisotropy (OMA) were treated by different models in order to establish a qualitative understanding of MAE origin~\cite{Bruno,andersson}. The OMA is given by the difference between the in-plane orbital magnetic moment $m_x$ and perpendicular orbital magnetic moment $m_z$, i.e $\Delta m=m_x-m_z$. To reveal the relationship between MAE and OMA, the total OMA ($\Delta m^{t}$) and the local OMA of Mn dopant ($\Delta m^{Mn}$) as a function of $U$ are presented in figure~(\ref{fig(cc0)}). In particular, the MAE can be related to the orbital magnetic moment through Bruno's formula~\cite{Bruno}, MAE$\displaystyle=\frac{\xi}{4\mu_B}\Delta m^{Mn}$. Here, the constant $\xi$ stands for the SOC strength. In other words, the magnetization easy axis coincides with the direction that has the largest orbital moment~\cite{Bruno}. For both Mn-MoS$_2$ and Mn-WS$_2$ systems, the coincidence between the magnetization easy axis and  the direction of largest orbital magnetic moment is obtained for each $U$ value. Indeed, by comparing MAE to the local OMA in figure~(\ref{fig(cc0)}), for a given $U$ value, the easy axis and the largest orbital moment are along the same direction. Regarding the change of local OMA upon the correlation effects, the sign reversal is consistent with that of MAE. However, despite the fact that the increase of $U$ enhances both the local OMA and MAE, their variations from $U=$ 2 to 4 eV are not similar. In particular, although the $U=$ 3 eV results in the largest MAE in Mn-MoS$_2$ and Mn-WS$_2$ systems, the local OMA associated with this value is not the largest among the rest of $U$ values.\\

Furthermore, the comparison between MAE and the total OMA which includes different contributions from the Mn environment, can be also tested. To this end, the total OMA versus the Hubbard parameter $U$ is plotted in figure~(\ref{fig(cc0)}). For the two Mn-MoS$_2$ and Mn-WS$_2$ systems, the correlation between MAE and the total OMA is shown by their sudden increase at $U=$ 1 eV. Unlike MAE, there is no sign reversal of the total OMA at $U=$ 1 eV in the case of Mn-WS$_2$ doped ML. This may due to the fact that the Bruno model is formulated for single atom but not for structures consisting of multiple atomic species with strong hybridization and large spin-orbit interaction~\cite{andersson}. However, this sign reversal, consistently with MAE, occurs for Mn-MoS$_2$ doped ML. \\

Overall, the MAE's enhancement is related to the OMA's increase for both the considered doped MLs. In particular, according to the table~(\ref{tab(c0)}),  the absolute values of the perpendicular orbital moment $|m_z^t|$ become considerably large when the correlation parameter increases from 0-1 eV to 2-4 eV. However, the in-plane orbital magnetic $|m_z^t|$ moments remain more or less constant. Therefore, the enhancement in  perpendicular magnetic anisotropy is mainly due to the enhancement of the perpendicular orbital magnetic moments imposed by the on-site electron correlation.\\
The behaviour of both MAE and OMA suggests the perpendicular orbital moment enhancement and the induced perpendicular magnetic anisotropy in the MLs have a common electronic origin. According to Refs.~\cite{kwon,pradhan,klatyk,huang}, for highly localized $3d$-orbitals including spin-orbit interaction and strong intra-atomic electron correlation, the orbital magnetic moment is anticipated to be large. In particular, this localization of the $3d$ orbitals allows atoms to retain their large atomic moments~\cite{kwon,pradhan,klatyk,huang}. However, strong hybridization leads to the broadening of the $3d$ bands producing much smaller orbital magnetic moments~\cite{kwon,pradhan,klatyk,huang}. As shown in figure~(\ref{fig:fig(c3)}), pDOS of $3d$ states of Mn dopant and $d$ and $p$ states of its neighboring M and X atoms, respectively, for $U=$ 0 eV and $U=$ 3 eV. In both cases, $U=$ 0 eV and $U=$ 3 eV, the high localization of Mn $3d$ states is obvious. The inclusion of $U$ pushes the $3d$ Mn localized states towards the Fermi level which weakens the overlapping between $3d$ Mn and its neighboring $3p$ an $4d$ orbitals of M and X host atoms. This effect enlarges the orbital magnetic moments and hence the MAE.\\

\subsubsection{Origin of magnetic anisotropy}

The spin-orbit coupling is responsible for both the  orbital moment and the magnetic anisotropy. The magnetization easy axis can be indicated by the competition between the second-order SOC energies per Mn dopants for in-plane ($E_{\text{SOC}}^{x}$) and out-of-plane ($E_{\text{SOC}}^{z}$) magnetization orientations, $\displaystyle \Delta E_{\text{SOC}}=E_{\text{SOC}}^{x}-E_{\text{SOC}}^{z}$~\cite{ref16,ref5}, where $E_{\text{SOC}}$ is given by the expression~(\ref{eqsoc}). As we can see in the figure~(\ref{fig(so)}), the sign of $\Delta E_{\text{SOC}}$ follows that of the MAE. In particular, apart from the case of $U=$ 0 eV and 1 eV, the out-of-plane spin-orbit energy is larger than the in-plane one which gives rise to the PMA. However, a different behavior between the variation of MAE and the variation of $\Delta E_{\text{SOC}}$ under $U$ is obtained as depicted in figure~(\ref{fig(so)}). In fact, these variations of $\Delta E_{\text{SOC}}$ are similar to those of the local OMA presented in figure~(\ref{fig(cc0)}). The discrepancy between MAE and $\Delta E_{\text{SOC}}$ variations under $U$-parameter can be understandable because a large SOC does not necessarily lead to large MAE~\cite{skomski}.\\

Three principal effects behind the magnetic anisotropy are the spin orbit effect, the crystal field effect and the exchange field effect~\cite{ref16,ref5}. All these effects are captured by the SOC second-order perturbation method~\cite{ref5} which is largely adopted to investigate the MAE origin. In fact, using the crystal field and the exchange splitting of Mn $3d$ orbitals together with the angular momentum operators $\hat{L}_{x}$ and $\hat{L}_z$, the MAE can be expressed as in Ref.~\cite{ref5},\\
\begin{equation} 
 \label{eq1}
 \text{MAE}\simeq\xi^2 \sum_{u^{\sigma},o^{\sigma'}}\frac{|\braket{u^{\sigma}|\hat{L}_z|o^{\sigma'}}|^2-|\braket{u^{\sigma}|\hat{L}_x|o^{\sigma'}}|^2}{\delta_{u^{\sigma},o^{\sigma'}}}(2\delta_{\sigma,\sigma'}-1)
 \end{equation}
 Here, the notations $\sigma,\sigma'=\uparrow,\downarrow$, indicate the spin directions of the occupied (o) and the unoccupied (u) states. $\delta_{u,o}=E_u-E_o$ is the energy difference between (o) and (u) states. The only non-zero matrix elements contributing to MAE are: $\braket{xz|\hat{L}_z|yz}=1$; $\braket{xy|\hat{L}_z|x^2-y^2}=2$;  $\braket{z^2|\hat{L}_x|xz,yz}=\sqrt{3}$; $\braket{xy|\hat{L}_x|xz,yz}=1$ et $\braket{x^2-y^2|\hat{L}_x|xz,yz}=1$. It is seen from Eq.~\ref{eq1} that the $\text{MAE}$ magnitude is governed by the weight of matrix elements and the energy differences $\delta_{u,o}$. Moreover, the sign of $\text{MAE}$ energy depends crucially on heavy matrix element, be it with positive or negative signs.\\
\begin{figure}
  \begin{center}
      \includegraphics[width=0.95\textwidth]{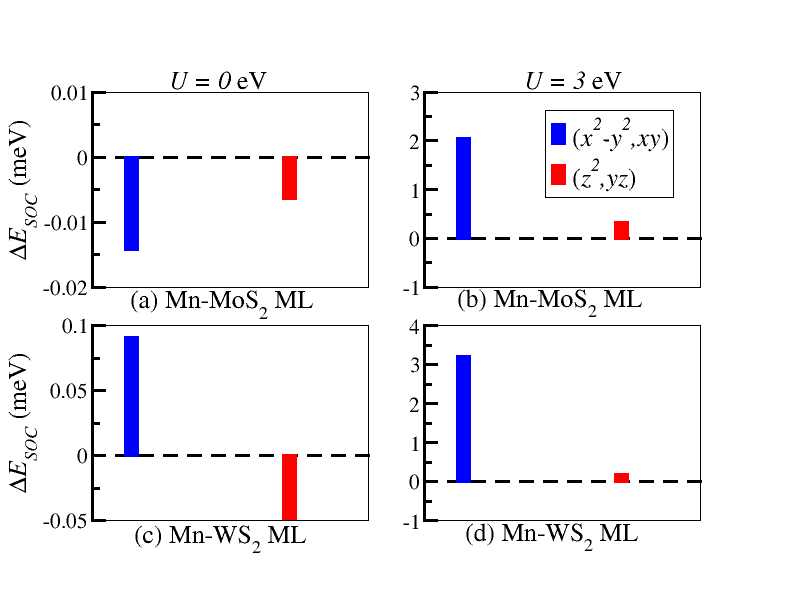} 
\end{center}
\caption{The DFT (a and c) and the DFT$+U$ (b and d) ($U=$ 3 eV) difference between the in-plane and out-of-plane $d$-orbital projected SOC energies of Mn dopant.
\label{fig:fig(c4)}}
\end{figure}
\begin{figure}
  \begin{center}
      \includegraphics[width=0.8\textwidth]{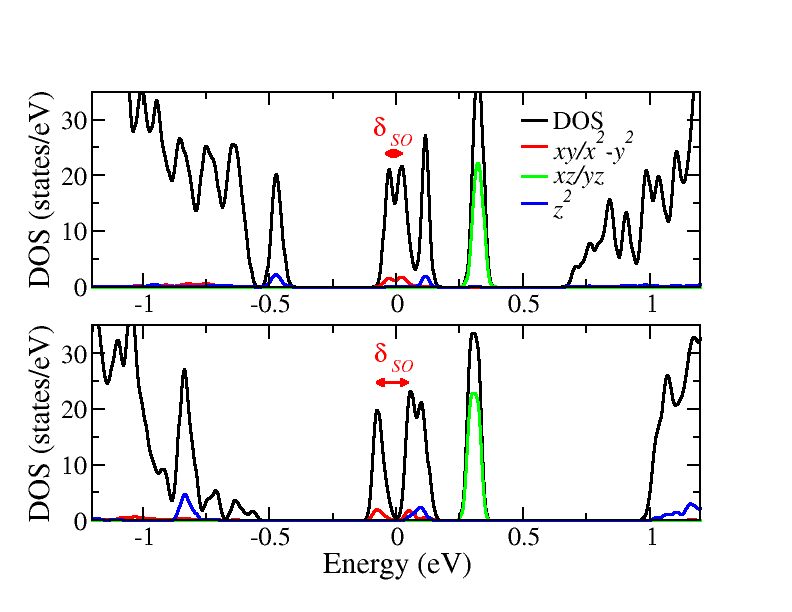} 
\end{center}
\caption{The DFT+$U$ DOS difference where the spin-orbit coupling is included for Mn-MoS$_2$ ML (upper panel) and Mn-WS$_2$ ML (down panel).
\label{fig:fig(ccp4)}}
\end{figure}
To understand the orbital origin responsible for the easy axis reversal, we consider two cases $U=$0 eV and $U=$ 3 eV. In particular, we analyse the pDOS of the \textit{3d} Mn orbitals to identify the nonvanishing matrix elements that contribute negatively or positively to MAE. For $U=$ 0 eV,  according to the figure~(\ref{fig(c31)}) of Mn-MoS$_2$ doped ML and the figure~(\ref{fig(c33)}) of Mn-WS$_2$ doped ML, the highest occupied states are spin up \textit{$d_{z^2}^{\uparrow}$} orbitals. According to Eq.~(\ref{eq1}), negative contribution to MAE comes from matrix elements, $\displaystyle \braket{{z^2}^{\uparrow}|\hat{L}_x|{xz}^{\uparrow},{yz}^{\uparrow}}$, while positive one corresponds to $ \displaystyle \braket{{z^2}^{\uparrow}|\hat{L}_x|{xz}^{\downarrow},{yz}^{\downarrow}}$. Those matrix elements have similar magnitudes. Owing to the intra-atomic Hund exchange splitting ($\Delta_{ex}$), recalling that spin-up and spin-down states with the same symmetry have different energies,~\cite{ref7} positive contributions of $\displaystyle \braket{{z^2}^{\uparrow}|\hat{L}_x|{xz}^{\uparrow},{yz}^{\uparrow}}$ remain small because of the large energy denominator between the two states. Hence, the $\text{MAE}$ is dominated by the negative contribution of $\displaystyle \braket{{z^2}^{\uparrow}|\hat{L}_x|{xz}^{\uparrow},{yz}^{\uparrow}}$.\\

For $U=$ 3 eV, according to the figure~(\ref{fig(c32)}) of Mn-MoS$_2$ doped ML and the figure~(\ref{fig(c34)}) of Mn-WS$_2$ doped ML, the band of the up-spin ${xy/x^2-y^2}^{\uparrow}$ is half occupied. This behavior dominates both the sign and value of the MAE by favoring the PMA. According to the expression~(\ref{eq1}), the dominant contribution is positive and stems from the non-zero matrix element $\braket{xy|\hat{L}_z|x^2-y^2}$ which is consistent with the DFT+$U$  calculated MAE. However, the degeneracy of the in-plane orbitals predict an infinite MAE which makes the second order perturbation method unreliable in this case. In Ref.~\cite{daalderop}, it was demonstrated that when the two orbitals $d_{xy}$ and $d_{x^2-y^2}$ are degenerate, and being at Fermi level, it favors the out-of-plane orientation magnetization. Indeed, the spin-orbit energy splitting reaches its maximum for these degenerate states when the magnetization is parallel to $z$-axis~\cite{daalderop}. Therefore, the PMA originates from the spin-orbit splitting of the degenerated Mn ${(d_{xy}^{\uparrow},d_{x^2-y^2}^{\uparrow})}$-orbitals placed at the Fermi level by means of the correlation effect.\\

Furthermore, using DFT and DFT+$U$, the decomposed $\Delta E_{\text{SOC}}$ on Mn 3\textit{d}-orbital, are presented in figure~(\ref{fig:fig(c4)}) for both Mn-MoS$_2$ and Mn-WS$_2$ doped MLs. The decomposed $\Delta E_{\text{SOC}}$ varies significantly by including the correlation effects. For the DFT case, the dominating contributions comes from the spin-orbit matrix elements involving either $d_{x^2-y^2}$ and $d_{xy}$. However, according to the pDOS, these coupled states are not occupied whereas only the SOC between the occupied and unoccupied states contributes to the magnetic anisotropy. The remaining large contribution comes from the coupling between $d_{z^2}$ and $d_{yz}$ orbitals which provides a negative contributions to $\Delta E_{\text{SOC}}$. Therefore, the negative MAE found for $U=$ 0 eV originates from the spin-orbit coupling between $d_{z^2}$ and $d_{yz}$ orbitals. This result is consistent with the second order perturbation theory method, modeled by Eq.~(\ref{eq1}), where the matrix element $\displaystyle \braket{{z^2}^{\uparrow}|\hat{L}_x|{xz}^{\uparrow},{yz}^{\uparrow}}$ is found dominant. For DFT+$U$, an important enhancement is observed for the magnitude of spin-orbit matrix elements involving $d_{x^2-y^2}$ and $d_{xy}$ states. These matrix elements are the major contribution to $\Delta E_{\text{SOC}}$. Together with the semi-metallic character of the in-plane orbitals, it is clear that they are behind the enhancement of MAE due to the correlations effect $U$.\\

The difference in MAE between the two MS$_2$ materials is related the behaviour of the in plane orbitals under spin-orbit coupling. The DOSs in figure~(\ref{fig:fig(ccp4)}) show that the spin-orbit splitting ($\delta_{SO}$) of ($d_{x^2-y^2}$,$d_{xy}$)-orbital energy is more important in Mn WS$_2$ system. This explains why Mn-WS$_2$ has larger MAE than that of Mn-MoS$_2$ ML. This observation means the the host TM material of Mn plays a critical role in determining the Mn induced anisotropy. In fact, for WS$_2$ ML, the spin-orbit coupling is stronger than for MoS$_2$ ML because W is heavier than Mo~\cite{Zhu}. In particular, the upper valence band of MS$_2$ around K point, showing strong spin-orbit splitting, is mainly composed by in-plane $p$-S and $d$-M orbitals~\cite{Zhu}. These latter hybridise with Mn in-plane orbitals leading to a stronger spin-orbit splitting  in the case of M=W.
\subsection{Effects of Mn-separation on MAE}
\begin{figure}[!h]
  \begin{center}
            \subfloat[\textbf{1 nn-z configuration}]{
      \includegraphics[width=0.35\textwidth]{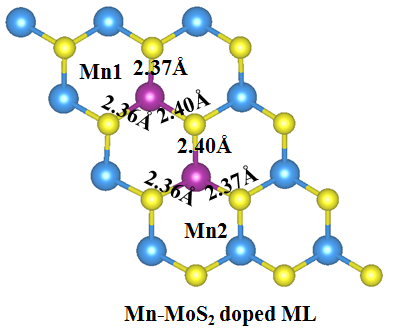}
      \includegraphics[width=0.35\textwidth]{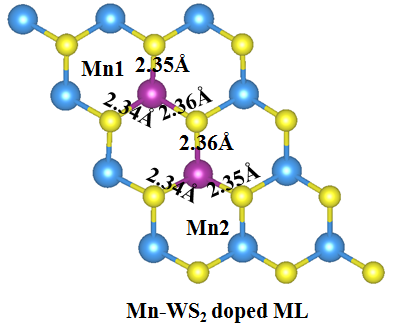}
      \label{fig(23)}
                         } 
     \qquad
    \subfloat[\textbf{2 nn-a configuration}]{
      \includegraphics[width=0.35\textwidth]{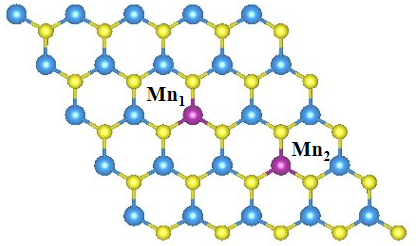}
      \label{fig(24)}
                         } 
    \subfloat[\textbf{3 nn-z configuration}]{
      \includegraphics[width=0.35\textwidth]{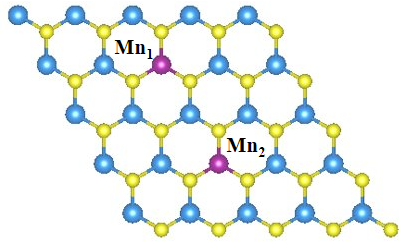}
      \label{fig(21)}
                         }
    \subfloat[\textbf{4 nn-a configuration}]{
      \includegraphics[width=0.35\textwidth]{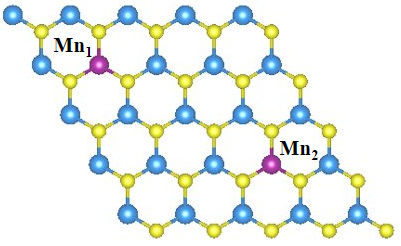}
      \label{fig(22)}
                         }                  
  \end{center}
\caption{Top view of the atomic structures  corresponding to 8\%-Mn-doped 5$\times$5$\times$1 MoS$_2$ supercells.  Four doping configurations have considered; (a) 1 NN configuration, (b) 2 NN configuration, (c) 3 NN configuration and (d) 4 NN configuration. The yellow atoms are the Sulfur (S), the blue atoms are the Molybdenum (Mo) and the Manganese atom is denoted by purple color.}
    \label{fig(2)}  
\end{figure}
\begin{figure*}
  \begin{center}
      \includegraphics[width=0.6\textwidth]{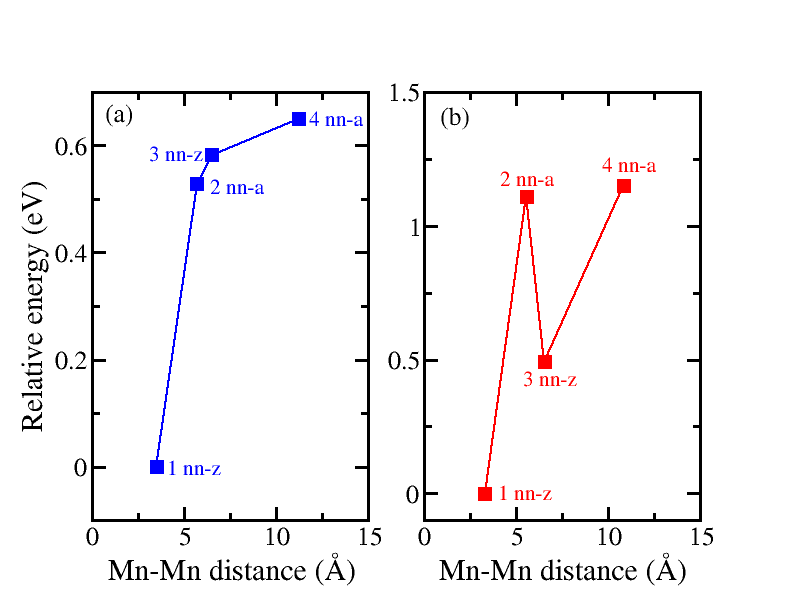}
    \caption{Mn-Mn distance dependence of relative energy of Mn pair doped (a) MoS$_2$ ML and (b) WS$_2$ ML.}
    \label{fig(c18)}
  \end{center}
\end{figure*}
\begin{figure*}
  \begin{center}
      \includegraphics[width=0.6\textwidth]{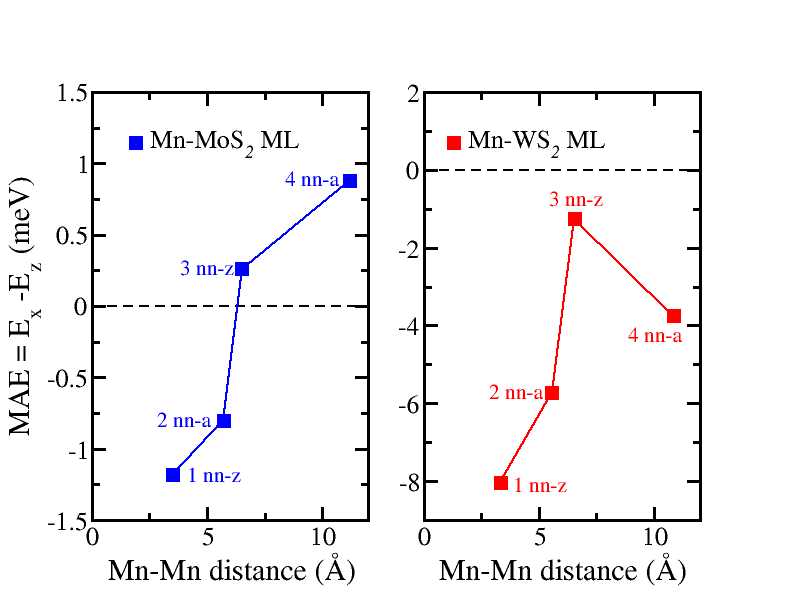}
    \caption{Mn-Mn distance dependence of (a) magnetic anisotropy energy and (b) the orbital magnetic anisotropy $\Delta m_0$ values for Mn-doped MoS$_2$.}
    \label{fig(c28)}
  \end{center}
\end{figure*}

\begin{figure*}
  \begin{center}
    \subfloat[\textbf{Mn MoS$_2$ ML}]{
      \includegraphics[width=0.65\textwidth]{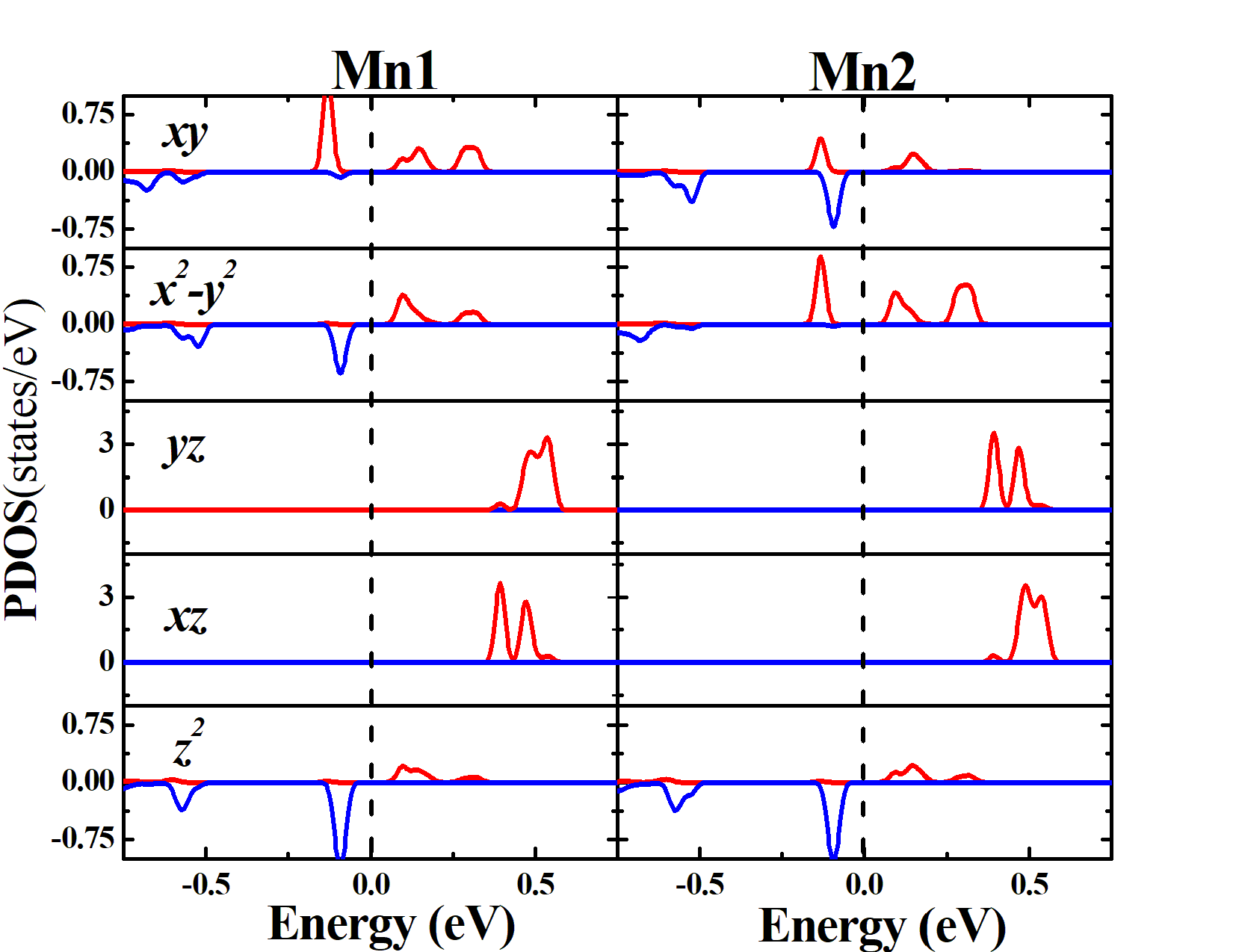}
      \label{fig(41)}
                         }
    \qquad
    \subfloat[\textbf{Mn WS$_2$ ML}]{
      \includegraphics[width=0.65\textwidth]{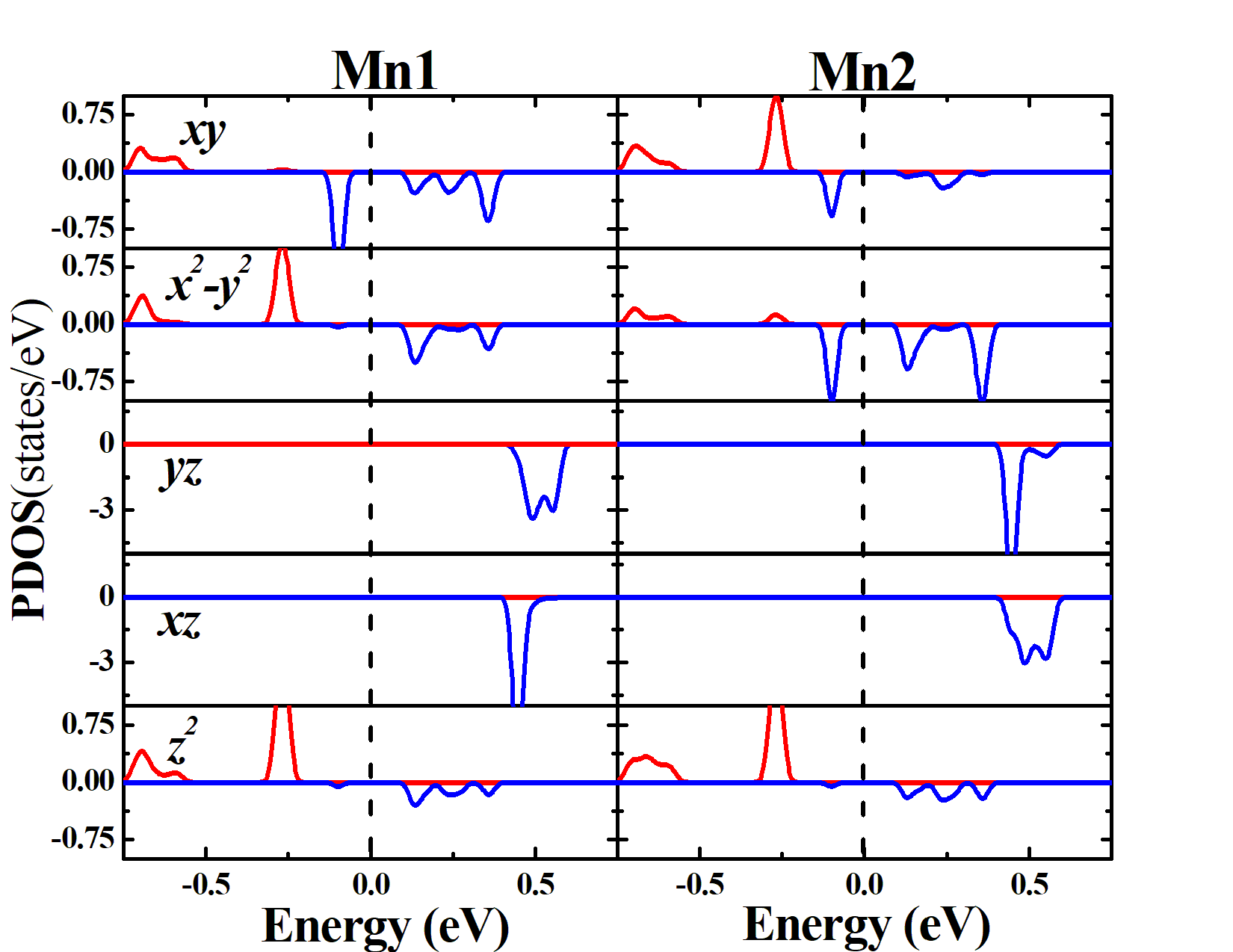}
      \label{fig(42)}
                         }
    \caption{The projected electron density of states of Mn$_1$ and Mn$_2$ $d$-orbitals in (a) MoS$_2$ ML and (b) WS$_2$ ML. The red and blue lines denote spin-up and spin-down channels, respectively. The Fermi level is indicated by the black dashed line.}
    \label{fig(c4)}
  \end{center}
\end{figure*}
In this section, we treat the case of Mn pairs implanted in a MoS$_2$ ML using 5$\times$5$\times$1 supercell (see figure~(\ref{fig(2)})). Depending on Mn-Mn distance, one can generate four different doping configurations: (i) first nearest neighbor (nn) configuration in which the two Mn impurities are in the nn positions of the zigzag chain (1 nn-z), (ii) the 2-nn configuration in which the two Mn impurities are in the second nn positions of the armchair chain (2 nn-a), (iii) the 3-nn configuration in which the two Mn impurities are in the third nn positions of the zigzag chain (3 nn-z) and (iv) the 4-nn configuration in which the two Mn impurities are in the nn positions of the armchair chain (4 nn-a).\\
\subsubsection{Structural stability}
We study the structural stability of the different configurations through the relative energy presented in figure~(\ref{fig(c18)}). The energy as a function of the Mn-Mn distance shows a behavior that differs from MoS$_2$ ML to WS$_2$ ML. In particular, for Mn-MoS$_2$ system, the relative energy increases with the increase of the dopant separation distance. Hence, the doped MoS$_2$ structure is more stable when the Mn impurities are close to each other, i.e 1 nn-z case. For Mn-WS$_2$ system, the relative energy does not show a monotone dependence on Mn-Mn distance. This may be due to the effect of the chain type (zigzag or armchair) on which the two dopants are placed. Same as Mn-MoS$_2$ system, in Mn-WS$_2$ doped ML, the lower relative energy is found for the doping configuration 1 nn-z. Therefore, in both systems, Mn-MS$_2$ MLs, Mn impurities prefer to be in the closest substituting M-sites.\\
\subsubsection{Magnetic anisotropy}
The magnetic anisotropy properties of Mn-MoS$_2$ and Mn-WS$_2$ systems are shown in figure~(\ref{fig(c28)}), in which MAEs are given for all doping configurations as a function of Mn-Mn distance. For the case of the host MoS$_2$, MAE increases and changes its sign with the increase of Mn-Mn distance. Hence, the magnetic anisotropy goes from IMA of -1.2 meV to PMA of about 1 meV when the Mn-Mn distance increases. In particular, it is found that 1 nn-z and 2 nn-a configurations have IMA, whereas the 3 nn-z and 4 nn-a configurations correspond to PMA. For WS$_2$ host material, MAE remains negative for all Mn-Mn distances with a minimum of -8 meV for 1 nn-z configuration. This latter corresponds to a large in-plane magnetic anisotropy. Besides for Mn WS$_2$ doped ML,  MAE is not monotonically depending on Mn-Mn distance. Although the Mn-Mn distances are roughly equal in the two configurations 2 nn-a and 3 nn-z, their corresponding MAEs are different. These results suggest that the type of atomic chain has an important role in determining MAE. Thus, we can say that for zigzag or armchair configurations, MAE increases with the increase of Mn-Mn distance. Overall, by comparing the MAEs obtained here and that of the single Mn doping cases, it is obvious that the diluted doping limit is required to enough the maintain a large PMA. 
\subsubsection{Origin of magnetic anisotropy}
In this section, as the Mn impurities atoms prefer to stay in the nearest positions at high concentrations, we choose the 1 nn-z MS$_2$ configurations for further study of magnetic anisotropy origin. The change of MAE in the case of the Mn-pairwise doping with respect to Mn-single doping can be related to the modification of Mn 3$d$ states. Indeed, putting another close dopant in the structure leads to the redistribution of 3$d$ states which is correlated with structural modifications around dopants. Therefore, we explore the structural and electronic feature changes induced by Mn-pairing in the 1 nn-z MS$_2$ configurations. The Mn-S bond lengths are given in figures~(\ref{fig(23)}). In the case of single dopant, the three Mn-S bonds are equal ($L$) obeying the $D_{3h}$ symmetry.
Contrary to the case of single dopant, which obeys the $D_{3h}$ symmetry; the lengths are no longer equal, indicating a symmetry reduction. The lengths of Mn$_1$-S and Mn$_2$-S bonds that relate the two Mn atoms are the largest lengths. This means that the two Mn dopant repel each other. In the meantime, the distances of Mn$_1$-S and Mn$_2$-S bonds, that relate each Mn atom to the nearest M of the zigzag chain, are the lowest lengths. To analyse the electronic structure, the pDOS are plotted in figures~(\ref{fig(41)}) and~(\ref{fig(42)}).  According to the figures~(\ref{fig(41)}) and~(\ref{fig(42)}), by placing the second Mn atom in a metal site close to the first one, induces electronic structure modifications. Considering in-plane orbitals, their degeneracy is lifted as shown in the figures~(\ref{fig(41)}) and~(\ref{fig(42)}). In fact, the Mn$_1$ $xy$-orbital is coupled to Mn$_2$ $d_{x^2-y^2}$-orbital as they are pointed toward each other because of the hexagonal geometry. The remaining Mn$_1$ $d_{x^2-y^2}$- and  Mn$_2$ $d_{xy}$-orbitals are mostly coupled to the $d$ orbital of the surrounding TM host atoms and they showing slightly different behaviour compared to each other. A quite similar effect is shown by out-of-plane orbitals  ($d_{xz}$, $d_{yz}$) where their associated degeneracy is lifted, as it can be seen in figures~(\ref{fig(41)}) and~(\ref{fig(42)}). In particular, the Mn$_1$ $d_{yz}$- and Mn$_2$ $d_{xz}$-orbital hybridize with the same 3$p$ of S cation while the Mn$_1$ $d_{xz}$ and Mn$_2$ $d_{yz}$ orbitals are coupled to S atom with Mo or W environment.\\

We now discuss MAE origin of Mn-MoS$_2$ doped ML, upon Mn-pairing. According to figure~(\ref{fig(41)}), the occupied states correspond to the orbitals $d_{xy}$, $d_{x^2-y^2}$ and $d_{z^2}$, while the unoccupied states correspond to $d_{xy}$, $d_{x^2-y^2}$, $d_{xz}$, and $d_{yz}$. For Mn$_1$ case, the prominent coupling elements are $\braket{{xy}^{\uparrow}|\hat{L}_z|{xz,yz}^{\uparrow}}$ and $\braket{{z^2}^{\downarrow}|\hat{L}_z|{xz,yz}^{\uparrow}}$. In particular, for Mn$_2$, the prominent coupling elements are $\braket{{x^2-y^2}^{\uparrow}|\hat{L}_z|{xz,yz}^{\uparrow}}$ and $\braket{{z^2}^{\downarrow}|\hat{L}_z|{xz,yz}^{\uparrow}}$. Here, $\braket{{xy}^{\uparrow}|\hat{L}_z|{xz,yz}^{\uparrow}}$ gives a positive contribution to MAE while $\braket{{z^2}^{\downarrow}|\hat{L}_z|{xz,yz}^{\uparrow}}$ gives a negative contribution. Because of the dominance of $\braket{{x^2-y^2}^{\uparrow}|\hat{L}_z|{xz,yz}^{\uparrow}}$ and $\braket{{xy}^{\uparrow}|\hat{L}_z|{xz,yz}^{\uparrow}}$, the 1 nn-z configuration of Mn-MoS$_2$ ML prefers IMA. In the case of Mn-MoS$_2$ doped ML, the matrix elements involving minority spin $d_{xy}^{\downarrow}$, $d_{x^2-y^2}^{\downarrow}$, $d_{xz}^{\downarrow}$ and $d_{yz}^{\downarrow}$ states yield positive contributions to MAE. In particular, for Mn$_1$, the positive contribution comes from $\braket{{xy}^{\downarrow}|\hat{L}_z|{xy,x^2-y^2}^{\downarrow}}$ and $\braket{{xy}^{\downarrow}|\hat{L}_z|{xz,yz}^{\downarrow}}$. Besides, for Mn$_1$, the positive contribution comes from $\braket{{x^2-y^2}^{\downarrow}|\hat{L}_z|{xy,x^2-y^2}^{\downarrow}}$ and $\braket{{x^2-y^2}^{\downarrow}|\hat{L}_z|{xz,yz}^{\downarrow}}$.  All these positive contributions finally lead to IMA for the 1 nn-z configuration of Mn-WS$_2$ ML.\\

 \section{Conclusion}
By performing DFT+U calculations, we have investigated the magnetic anisotropy induced by Mn doping in MoS$_2$ and WS$_2$ MLs. In the case of a well-isolated Mn atom substituting a W center, a large positive MAE of 35 meV is obtained for Mn-WS$_2$ doped ML. For Mn-MoS$_2$ doped ML a smaller MAE of 8 meV is found. These results mean that the preferential direction of magnetization is perpendicular to the ML plane when the origin of large MAE is attributed to the presence of in-plane Mn orbitals located in the vicinity of the Fermi level. In the case of the Mn pairwise doping, a spin reorientation transition from out-of-plane to in-plane magnetization takes place when the Mn-Mn distance decreases. In general, it is apparent that the clustering of Mn dopants favors the in-plane magnetization. Our findings show that important magnetic anisotropy can be found in Mn-doped WS$_2$ and MoS$_2$ ML only for considerable Mn-Mn distances.\\

\section*{Acknowledgments}
S.L. acknowledges funding from the Priority Programme SPP 2244 2D Materials - Physics of van der Waals Heterostructures of the Deutsche Forschungsgemeinschaft (DFG) (project LO 1659/7-1) and the European Research
Council (ERC) under the European Union's Horizon 2020 research and
innovation programme (ERC-consolidator Grant No. 681405 DYNASORE).
I.C.G. thanks the CALMIP initiative for the generous allocation of computational time, through the project p0812, as well as GENCI-CINES and GENCI-IDRIS for grant A008096649.  
\bibliographystyle{apsrev}
\bibliography{adlen_12}
\end{document}